\documentclass[aps,prl,twocolumn,superscriptaddress,showpacs]{revtex4}
\usepackage{amsmath}
\usepackage{graphicx}% Include figure files

\begin{document}

\title{A model for coexistent superconductivity and
ferromagnetism}

\author{Jason Jackiewicz}
\affiliation{Department of Physics, Boston College, Chestnut Hill, MA, 02467}

\author{Krastan B. Blagoev}
\affiliation{Theoretical Division, Los Alamos National Laboratory, Los Alamos, NM, 87545}

\author{Kevin S. Bedell}
\affiliation{Department of Physics, Boston College, Chestnut Hill, MA, 02467}

\date{\today}

\begin{abstract}
We explore the various temperature dependencies and thermodynamic quantities of a mean
field model
of a ferromagnetic-superconducting system.  The starting point for this model
is based on an s-wave pairing scheme
in the singlet channel of the superconducting state and a spontaneously broken symmetry
phase in the ferromagnetic state.  We show numerically
and analytically that a state of coexistence reveals itself and is favored energetically
over other possible states, and a simple phase diagram is developed. Finally, a comparison
of the specific heat with experiment is shown.
\end{abstract}


\maketitle

%\section{\label{sect-intro}Introduction}

Much work in the past several decades has been done studying the
interplay between magnetism and superconductivity.  Some of
the early progress by Abrikosov \cite{abrikosov1988}, and the work leading to the
so-called LOFF model \cite{larkin1964,fulde1964} was
directed towards ferromagnetic systems
of magnetic impurities coexisting with superconducting order.
Fay and Appel \cite{fay1980} showed how itinerant ferromagnetism can
promote p-wave superconductivity, and much of the theoretical
studies was concentrated on developing the symmetry of the
superconducting order parameter.
The question of possible s-wave singlet pairing was thought possible
on the paramagnetic side of the magnetic transition, but it was shown even
there that ferromagnetic fluctuations destroy it near the Curie
temperature \cite{berk1966}.

More recently, two of the current authors studied a weak ferromagnetic
Fermi liquid and showed that s-wave superconductivity is possible, and
that it is the favored phase on the \textit{ferromagnetic}
side \cite{blagoev1998,blagoev1999}.  A possible phase diagram was
predicted, however, at the time there were no materials that
demonstrated superconductivity inside the ferromagnetic state.
Recently a similar consideration of s-wave superconductivity has been carried out
by Suhl \cite{suhl2001} and Abrikosov \cite{abrikosov2001} in which the ferromagnetism
is due to localized spins.
We instead develop an itinerant ferromagnetic model in which the magnetic electrons are also the ones
responsible for the formation of the Cooper pairs.
In this work, it must be noted that the BCS state is actually the preferred one
energetically, as shown in ref.\cite{lei1984}.  However, we are assuming
that there \textit{already exists} weak ferromagnetic order, inside which arises superconductivity.
The point is that we are assuming the strongest competition to be
between the coexistent and the ferromagnetic state, with the pure BCS state not
being a possibility.  Moreover, in the microscopic models discussed
 \cite{blagoev1998,blagoev1999,suhl2001,abrikosov2001}, the s-wave superconductivity
pairing interaction vanishes when the ferromagnetic order vanishes.
With that said, we can consider this an effective mean field
that has nontrivial results and can be shown to coincide with what is seen experimentally.

The phenomenon of coexistent superconductivity and ferromagnetism
was then discovered in UGe$_{2}$ \cite{saxena2000,bauer2001}, and subsequently
in ZrZn$_{2}$ \cite{pfleiderer2001} and URhGe \cite{aoki2001}.  The phase diagram
of UGe$_{2}$ is very similar to the one proposed in
refs.\cite{blagoev1998,blagoev1999}, however,
no superconductivity was detected in any of these materials on the
paramagnetic side, contrary to early predictions.  These currrent discoveries
are promoting much more theoretical
work to explain the superconducting pairing, with most calculations based on a
p-wave triplet state \cite{wang2001}. Also, it has been verified \cite{pfleiderer2002}
that the magnetic
transition in at least one of these materials, UGe$_{2}$, is of first-order below
a certain temperature.  We will not discuss the thermodynamics of that transition
here, as it will be reserved for a future publication.

In this paper we will present some new and unexpected results of an alternative
model of s-wave singlet pairing developed by Karchev et al.\cite{karchev2001}.
In that paper, some zero temperature results were calculated, and the finite temperature
results will be shown here.

%\section{\label{sect-model} The Model}

We refer the reader to ref.\cite{karchev2001} for the details, and here
we look at the mean-field
Hamiltonian obtained from a model Hamiltonian by the standard mean-field procedure,
\begin{eqnarray}
\label{mf-ham}
\nonumber
H_{mf}&=&\sum_{\vec{p}}\epsilon_{p} (c_{\vec{p}}{}_{\uparrow}^{\dag}
c_{\vec{p}}{}_{\uparrow}+ c_{\vec{p}}{}_{\downarrow}^{\dag}c_{\vec{p}}{}_{\downarrow})\\
\nonumber
      &+&\frac{JM}{2}\sum_{\vec{p}}(c_{\vec{p}}{}_{\uparrow}^{\dag}
c_{\vec{p}}{}_{\uparrow}- c_{\vec{p}}{}_{\downarrow}^{\dag}c_{\vec{p}}{}_{\downarrow})\\
      &-&\sum_{\vec{p}}(\Delta c_{\vec{p}}{}_{\uparrow}^{\dag}c_{-\vec{p}}
{}_{\downarrow}^{\dag}+H.c.)+\frac{1}{2}JM^{2}+\frac{|\Delta|^{2}}{g}.
\end{eqnarray}

\noindent  The diagonalization of this Hamiltonian using a
Bogoliubov transformation yields,
\begin{equation}
H_{MF}=E_{0} + \sum_ {\vec{p}}\left(E_{p}^{\alpha}\alpha _{\vec{p}}^{\dagger}
\alpha _{\vec{p}}+E_{p}^{\beta}\beta _{\vec{p}}^{\dagger}
\beta _{\vec{p}}\right),
\end{equation}
\noindent where
\begin{eqnarray}
\nonumber
E_{0} = \sum_{\vec{p}}\epsilon_{\vec{p}}^{\downarrow} +
\frac{1}{2}JM^{2}+\frac{|\Delta|^{2}}{g},  \\
\nonumber
\epsilon_{\vec{p}}^{\downarrow}{}^{\uparrow} = \frac{p^{2}}{2m^{*}} -
\mu \mp \frac{JM}{2}.
\end{eqnarray}

\noindent The quasiparticle energy dispersion relations are,
\begin{eqnarray}
\label{dispersion}
E_{p}^{\alpha} = \frac{JM}{2} + \sqrt{\xi_{p}^{2} + |\Delta|^{2}}, \\
E_{p}^{\beta}  = \frac{JM}{2} - \sqrt{\xi_{p}^{2} + |\Delta|^{2}}.
\end{eqnarray}

\noindent The final step is to minimize the free energy to produce the
mean-field equations.  This results in a set of two coupled equations in
$M$ and $\Delta$ that will be solved self-consistently below. For $M$ we find,
\begin{equation}
\label{total-mag}
M=\frac{1}{2}\int\frac{d^{3}p}{(2\pi)^3}(1-n_{p}^{\alpha}-n_{p}^{\beta}),
\end{equation}
\noindent and for $\Delta$,
\begin{equation}
\label{total-gap}
|\Delta|=\frac{|\Delta| g}{2}\int\frac{d^{3}p}{(2\pi)^3}\frac{n_{p}^{\beta}
-n_{p}^{\alpha}}{\sqrt{\xi_{p}^{2} + |\Delta|^{2}}}.
\end{equation}

%another way of splitting equations
%\begin{equation}
%\begin{split}
%H_{mf} &=\sum_{\vec{p}}\epsilon_{p} (c_{\vec{p}}{}_{\uparrow}^{\dag}
%c_{\vec{p}}{}_{\uparrow}+ c_{\vec{p}}{}_{\downarrow}^{\dag}c_{\vec{p}}{}_{\downarrow})\\
%       &+\frac{JM}{2}\sum_{\vec{p}}(c_{\vec{p}}{}_{\uparrow}^{\dag}c_{\vec{p}}
%{}_{\uparrow}- c_{\vec{p}}{}_{\downarrow}^{\dag}c_{\vec{p}}{}_{\downarrow})\\
%       &-\sum_{\vec{p}}(\Delta c_{\vec{p}}{}_{\uparrow}^{\dag}c_{-\vec{p}}{}_{\downarrow}^{\dag}
%         +H.c.) + \frac{1}{2}JM^{2}+\frac{|\Delta|^{2}}{g}
%\end{split}
%\end{equation}

%\section{\label{sect-zerotemp} Zero Temperature}
The two order parameters, $M$ and $\Delta$, have
dependencies such as $M=M(g,J)$ and $\Delta=\Delta(g,J)$ at $T=0$.  From the numerical solutions
of the two coupled equations (\ref{total-mag},\ref{total-gap}),
 we show characteristic curves of the order parameters,
a phase diagram, and the free energy difference of different states.  We stress that all
of the future results are derived strictly from the dispersion relations and the mean-field
equations only, with no other assumptions made about coupling strength limits or small
magnetization.

At $T=0$, only one of the fermionic particles contributes, since $E_{p}^{\alpha}>0$
for all $p$, but $E_{p}^{\beta}<0$ when $p<p_{F}^{-}$ and $p>p_{F}^{+}$, where
\begin{equation}
p_{F}^{\pm}=\sqrt{2m^{*}\mu \pm m^{*}\sqrt{(JM)^{2}-4|\Delta|^{2}}}.
\label{pf-pm}
\end{equation}

\noindent Therefore, the coupled equations (\ref{total-mag},\ref{total-gap}) reduce to
\begin{equation}
M = \frac{1}{12\pi^{2}} \left[ (p_{F}^{+})^{3} - (p_{F}^{-})^{3}\right],
\label{coup-m-0}
\end{equation}
\noindent and
\begin{equation}
\begin{split}
|\Delta| = \frac{g|\Delta|}{2}[&\int_{0}^{\infty}{\frac{d^{3}p}{(2\pi)^{3}}}
 {\frac{1}{\sqrt{\xi_{p}^{2} + |\Delta|^{2}}}}\\
                                     &- \int_{p_{F}^{-}}^{p_{F}^{+}}
 {\frac{d^{3}p}{(2\pi)^{3}}}
 {\frac{1}{\sqrt{\xi_{p}^{2} + |\Delta|^{2}}}}].
\label{coup-gap-0}
\end{split}
\end{equation}

\noindent The only solutions of eqs.(\ref{coup-m-0}) and (\ref{coup-gap-0}) which
exhibit the coexistence of superconductivity and ferromagnetism are in the case
when $JM>2|\Delta|$.  However, in theory, there are two other solutions: the pure
ferromagnetic state with $\Delta=0$ and the BCS state with $M=0$ (and obviously
the normal, paramagnetic state).  We will not consider
the 'BCS' state here (see introduction).
%since from eq.(\ref{pf-pm}) there is an imaginary part when $2|\Delta|>JM$.
%This can be shown to be corrected in the complete finite temperature equations
%\ref{total-mag,total-gap}
%in the limit of $T\rightarrow 0$ where a 'BCS' state occurs without problems.

Our concern will be with the case \mbox{$JM>2|\Delta|$}
as in ref.\cite{karchev2001}. A note is in order here concerning the approximations made in
refs.\cite{karchev2001,shen2002}.  Analytically, the solutions
obtained from the coupled equations at $T=0$
when $JM$ is close to, but larger than $2\Delta$ are
\begin{equation}
\label{coup-approx}
I=\frac{r}{\sqrt{r^{2}-1}}\Delta ,
\;\;\;
\Delta =\sqrt{\frac{r-1}{r+1}}\Delta _{0},
\end{equation}
\noindent where $\Delta _{0}$ is the usual energy gap equation in a non-magnetic
superconducting state:
 \begin{equation}
\label{}
\Delta _{0}=2\epsilon _{c} \exp \left(-\frac{1}{gN(0)}\right).
\end{equation}

\begin{figure}
%\vskip 0.15cm
\includegraphics[width=8.5 cm,height=5cm]{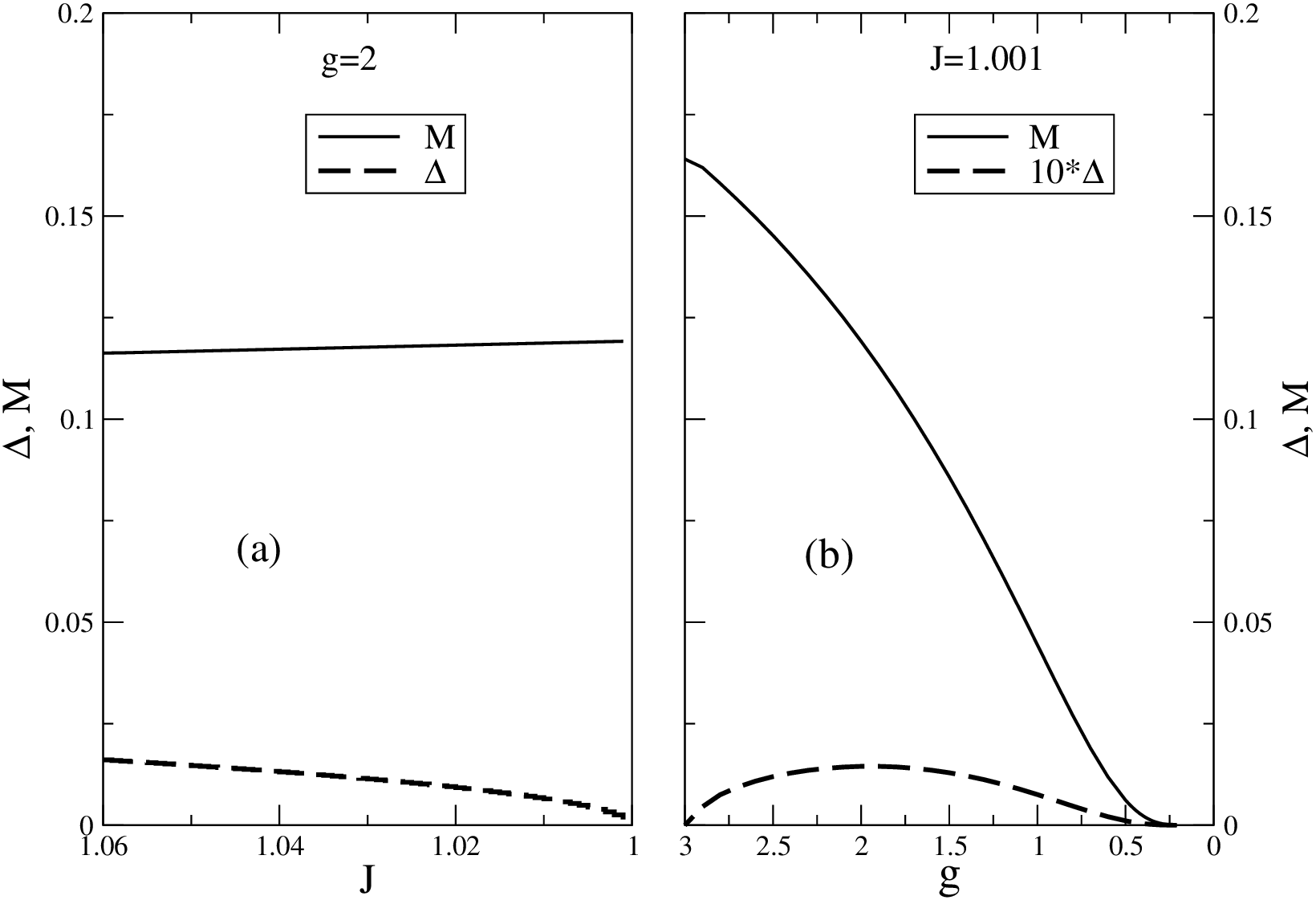}
\caption{M and $\Delta$ curves at $T=0$ as functions of $J$ and $g$.  In (a) we see
that these order parameters are weak functions of $J$, while in (b)
there is strong $g$-dependence. Note: Here and elsewhere, by $g$ we mean
$2\pi^{2}gN(0)$, so that these calculations are definitely in the weak-coupling
regime($gN(0)<<1$).  Also, the magnetization is normalized by $n$, so that $M/n\leq 1$.}
\label{m-delta-versus-g-J-graph}
\end{figure}

\noindent The important thing to notice is that in eq.(\ref{coup-approx}) the magnetization and
the gap are strictly proportional to each other in the coexistent state.
 This is an artifact of the
approximation used, and we would like to stress that the numerical calculations
do not make use of this approximation.  To show this explicitly, we look at
Fig.\ref{m-delta-versus-g-J-graph} where the coexistent solution with both order
parameters is shown as a function of $J$ and $g$ respectively.

It is seen from this figure that both order parameters go to zero together either at
$J=1$ or $g=0$.  This is due to the self-consistency of the coupled equations. As $J$
is increased the order parameters are fairly constant and solutions can be found for
large values of $J$.
However, we are mainly interested in weak ferromagnetism, i.e.,
when the coupling $J$ is close to 1, and we take the value of $J=1.001$.  Using this $J$ we
show the order parameters as functions of $g$ in (b) of Fig.\ref{m-delta-versus-g-J-graph}.
This has an interesting feature in that the superconducting gap increases from $g=0$
and then gets suppressed at a higher value of $g$.  The
magnetization is constant after the gap disappears because $M$ is not an
explicit function of $g$, only implicitly through $\Delta$.  This feature of
the gap increasing and then being suppressed is non-existent in the approximate
eqns.(\ref{coup-approx}).
 %This will have important effects on the free
%energy as will be seen below.

%The diagram in Fig.\ref{m-delta-versus-g-J-graph}b looks qualitatively similar to
%the experimental phase diagram
%of $UGe_{2}$, except here, $T=0$.  The difficulty arises in
%identifying an exact 'pressure' axis in terms of coupling constants,
%and for the time being we will use $g$.

\begin{figure} [b]
\smallskip
\includegraphics[width=8.5 cm,height=5cm]{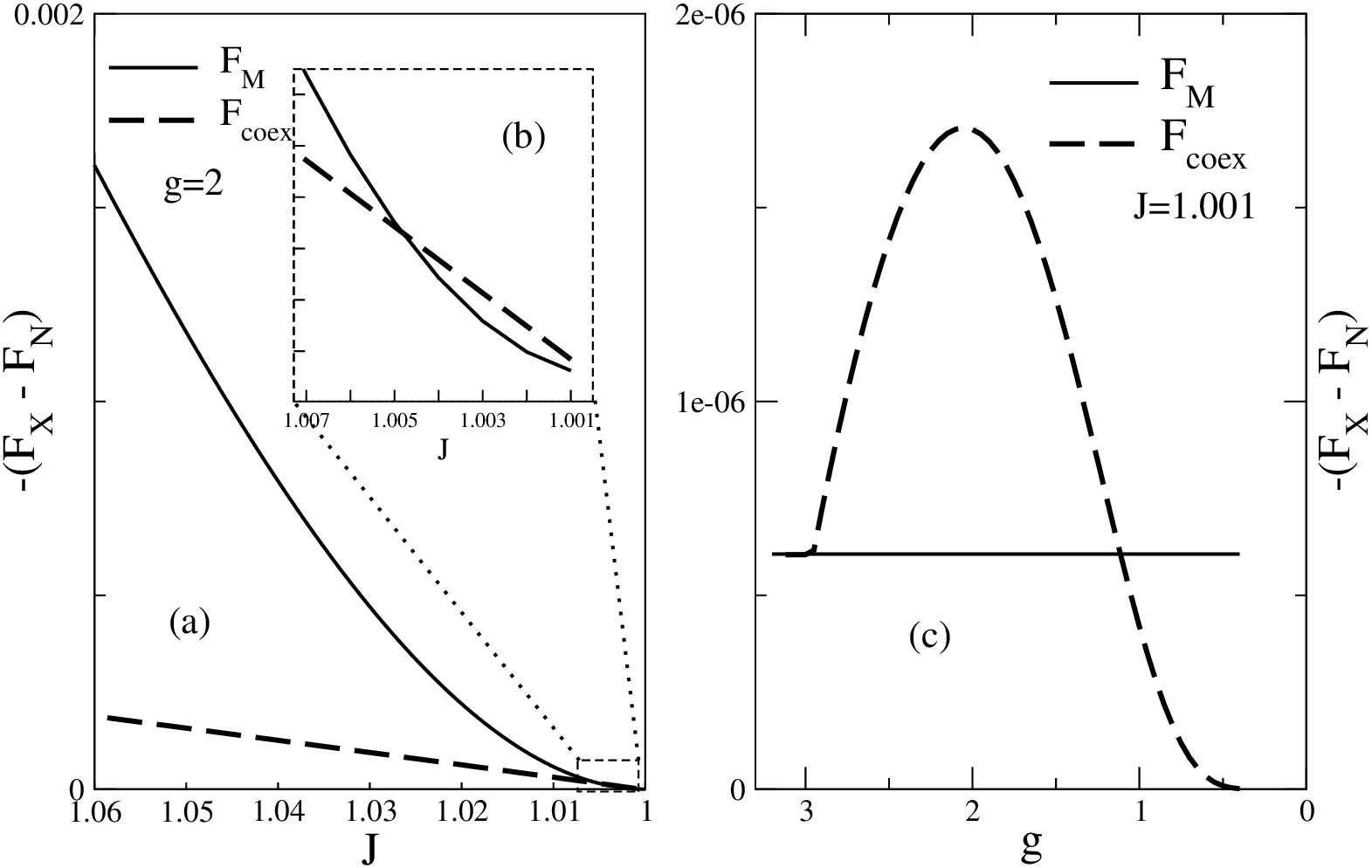}
\caption{Free energy of various states with respect to the normal
state($\Delta=M=0$).  Notice where the coexistent
state has its lowest free energy value (see Fig.\ref{m-delta-versus-g-J-graph}),
i.e., where the gap has its largest value.}
\label{free-energy-T=0-graph}
\end{figure}

In Fig.\ref{free-energy-T=0-graph} we have calculated the free energy of the coexistent state
with relation to the normal state and the pure ferromagnetic state ($\Delta=0$).
 In the
range where the superconducting gap has its largest value, $g\approx2$
(see Fig.\ref{m-delta-versus-g-J-graph}b), the free energy of the superconducting
ferromagnetic state is lower than the two other states, and hence is the
preferred one.  For very small values of g, we see a crossover.  This behavior
will also be evident in the finite temperature free energy, where the area
around the largest value for $T_{c}$ in the superconducting state yields a
lower free energy.

%\section{\label{sect-finitetemp} Finite Temperature}

At finite temperatures one can look at the same types of dependencies
of the order parameters.  We solve eqtns.(\ref{total-mag},\ref{total-gap}) introducing
a debye cutoff in the otherwise divergent integrals.  The cutoff is small compared to the
Fermi energy as this is the region where the important physics takes place, and varying
this cutoff has no qualitative effects on the various quantities calculated below.
A phase diagram can also be developed which
shows how the Tc's depend on the coupling constants. Then,
thermodynamic properties can be calculated such as the free energy and specific heat.

\begin{figure}
\includegraphics [width=8.5 cm,height=5cm]{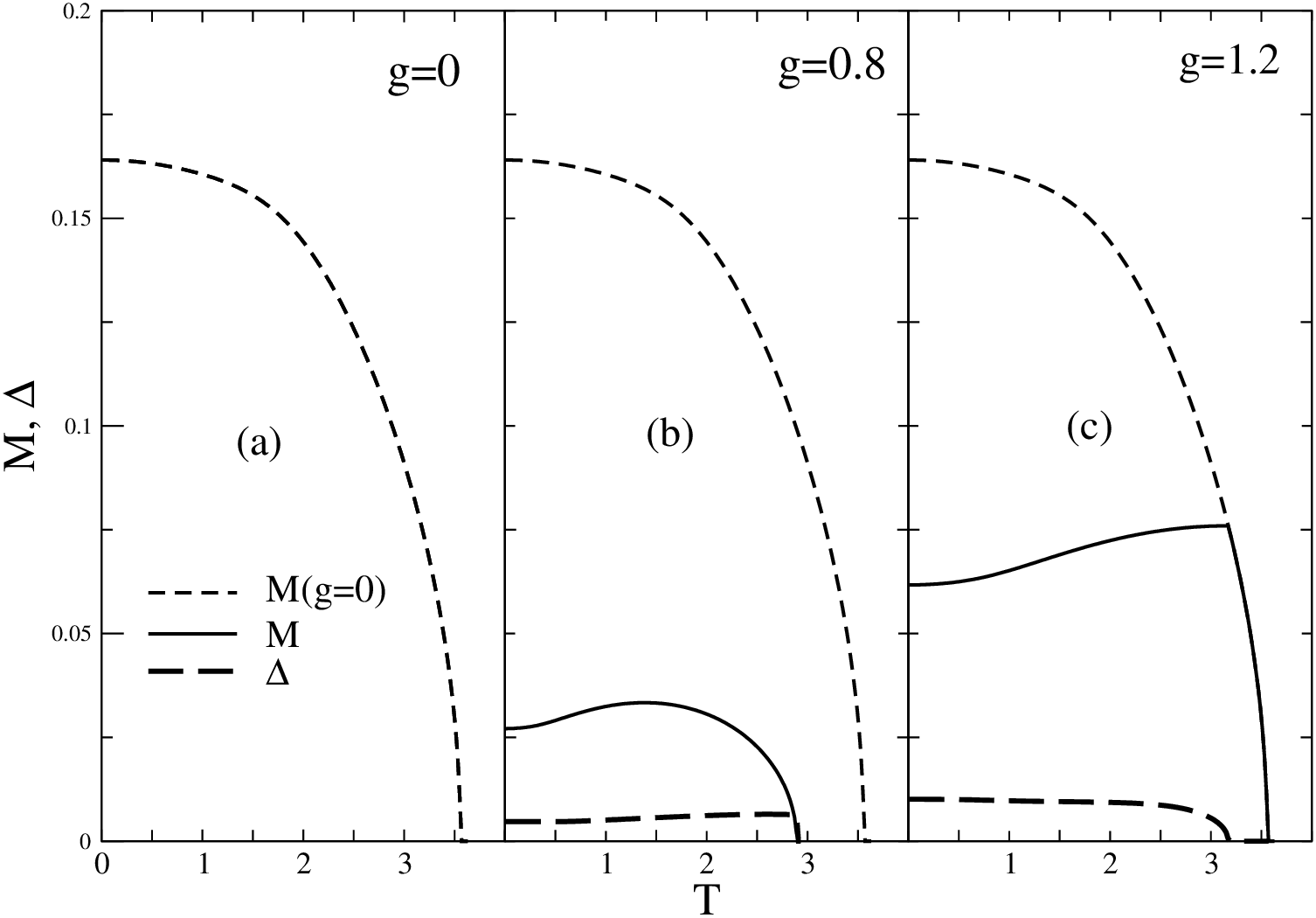}
\caption{The temperature dependence of the order parameters
for different coupling values of $g$. Here, $J=1.001$. The temperature, $T$, is unitless, but
can be converted to Kelvin with one input parameter, the Fermi energy.  For example, using
this conversion for $UGe_2$, the approximate temperature is $\approx5T$[K].  This gives
a Curie temperature of the right order of magnitude.}
\label{T-dependence-graph}
\end{figure}

The temperature dependence of $M$ and $\Delta$ can differ widely with respect
to changes in the coupling constant space.  This is why this model is convenient
because it is adaptable for different cases.  By tuning the parameters we can
see physically realizable results.  Figure \ref{T-dependence-graph} shows some examples.
In (a), $g=0$ and we see the full magnetization as it approaches its Curie temperature.
There is no superconducting gap in this case.  In the second and third graphs, we now turn
on the superconducting coupling and see that it has dramatic effects on the
magnetization.  There are two cases.  For a small $g$, the $T_{c}$ for $M$
and $\Delta$ coincide.  The superconductivity weakens the magnetization and they
become proportional and simultaneously go to zero.  The second case is for a larger $g$
where now the $T_{c}$'s differ.  The gap disappears and the magnetization falls back
onto the full saturation curve, as expected from the equations.

%\subsection{\label{phase-diagram} Phase Diagram}

\begin{figure}
\includegraphics[width=8.5 cm,height=5cm]{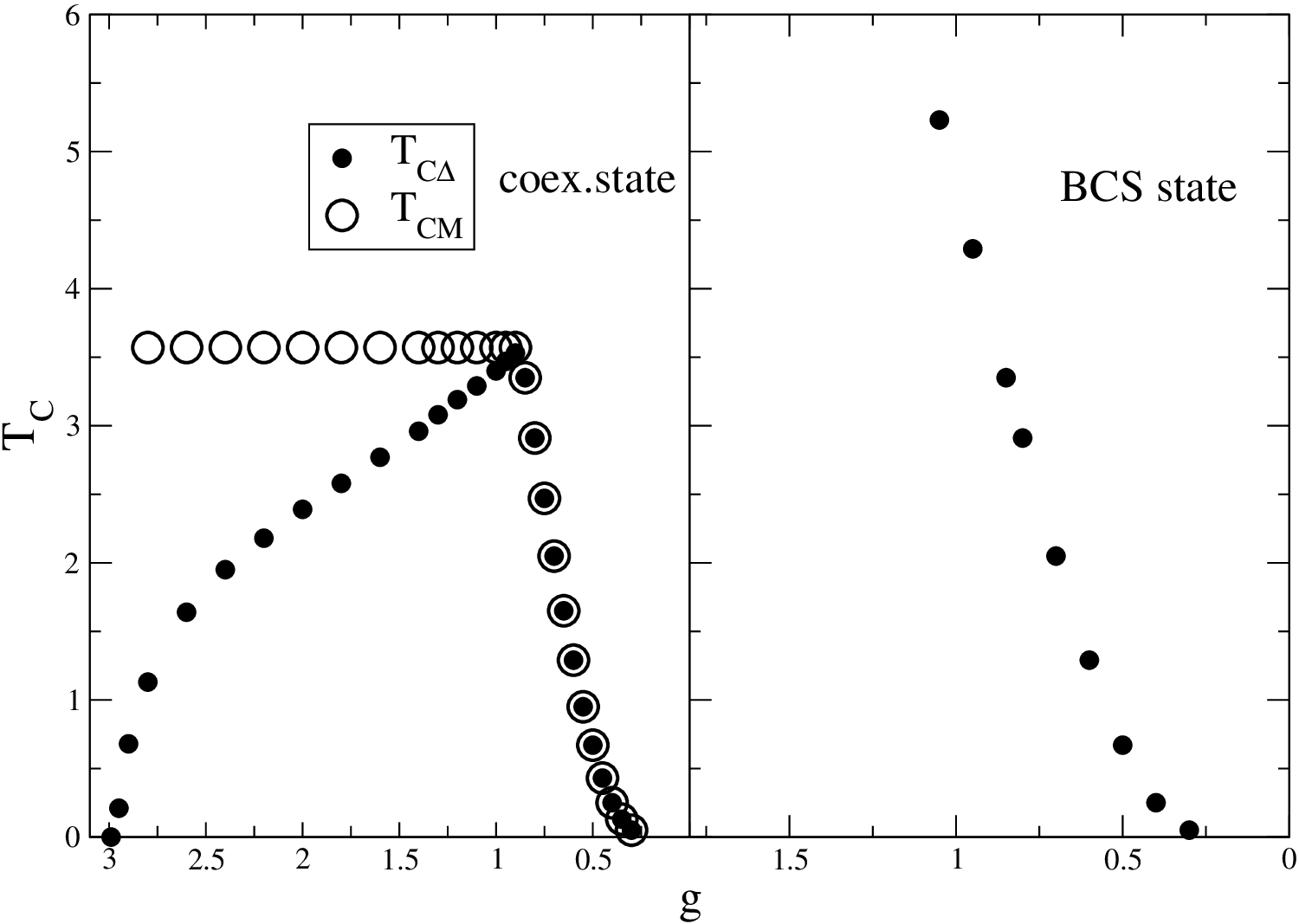}
\caption{A typical phase diagram for a fixed value of J=1.001 with
the $T_{c}$'s on the y-axis.  Also shown for comparison is the $T_{c}$ dependence of a pure BCS
superconductor on $g$.}
\label{phase-diagram-graph}
\end{figure}

In Fig.\ref{phase-diagram-graph} a phase diagram is developed from this model.
It shows the $T_{c}$'s of each order parameter of the coexistent
solution plotted as functions of $g$.  In (b) is
the phase diagram of a normal BCS superconductor without magnetic ordering
and in comparison with (a), it is obvious that the effect of the
magnetization as it gets large is to suppress and then kill superconductivity.
The superconducting $T_{c}$ goes through a maximum and then is suppressed eventually to zero.

%\subsection{\label{free-energy} Free Energy}

\begin{figure}
%\vskip 1.0cm
\bigskip
\includegraphics[width=8.5 cm,height=5.0cm]{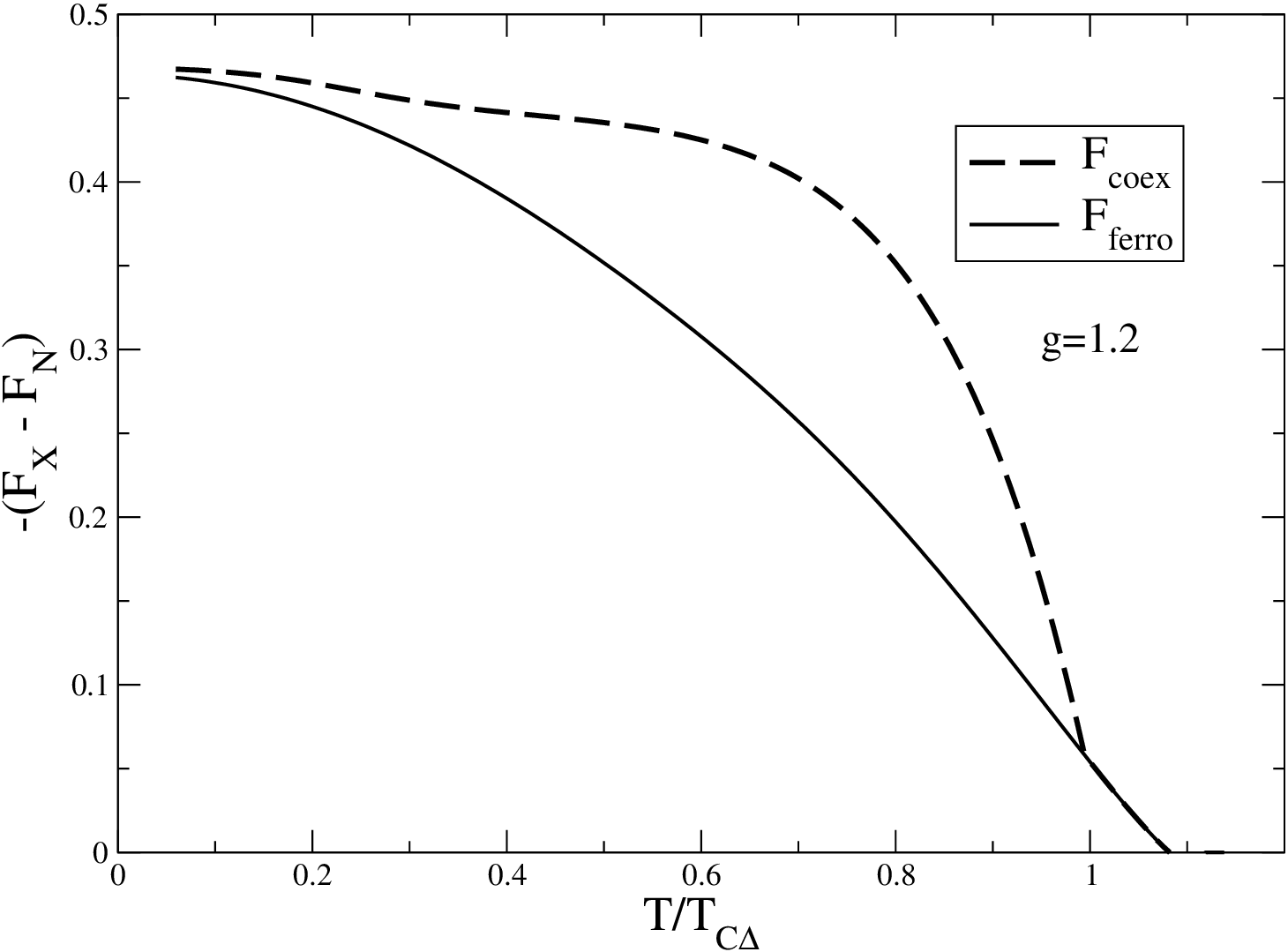}
\caption{The y-axis is labelled as the negative of the free energy difference of a certain
state and the normal state, in arbitrary units. $T=T_{c\Delta}$ where the superconducting gap
parameter vanishes inside the magnetic state, at the point where the
coexistant$\rightarrow$ferromagnetic transition is.  The dashed curve does merge smoothly
yet abruptly with the solid curve, at which point the two lines become indistinguishable.}
\label{nonzeroT-freeEn-graph}
\end{figure}

As with the zero temperature case, the free energy of the coexistent state is lower than
that of the pure ferromagnetic state through a broad temperature range for certain
values of the coupling constants.  In Fig.\ref{nonzeroT-freeEn-graph} we show one case.
This particular plot is calculated with $g=1.2$, hence it is helpful to refer to
Fig.\ref{phase-diagram-graph}(a) to see which part of the phase diagram we are analyzing.
In this case, we are looking at the zone where the $T_{c}$'s of the two order parameters
do not coincide.

%\subsection{\label{spec-heat} Specific Heat}

\begin{figure}
\bigskip
\includegraphics[width=7.5 cm,height=5cm]{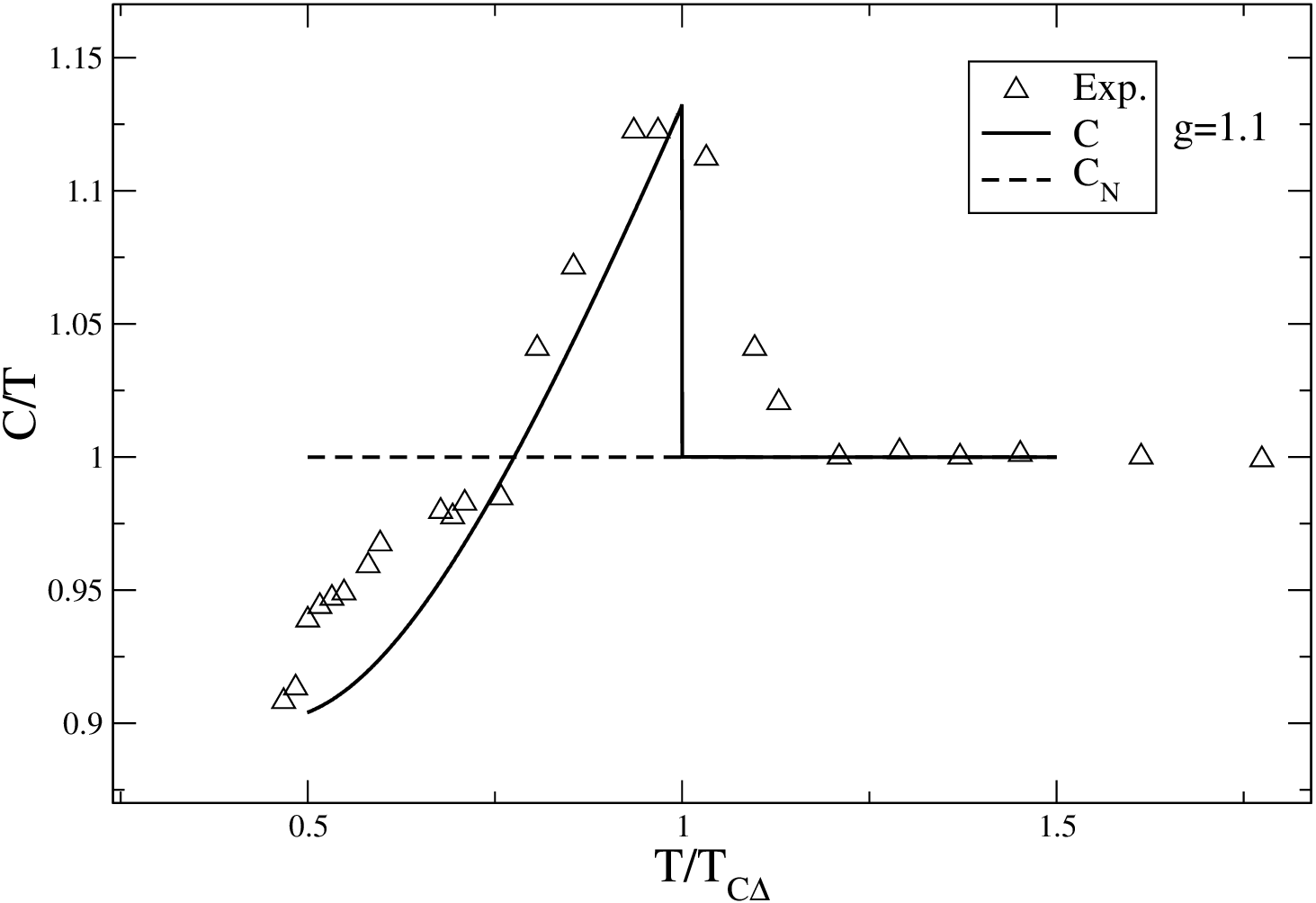}
\caption{The specific heat jump is much smaller than
the BCS one, in this case about 15 percent. The (rescaled)experimental data
is taken from ref.\cite{tateiwa2001}}
\label{spec-heat-graph}
\end{figure}

Experiments have indicated that the specific heat displays
a very small jump ($UGe_{2}$)\cite{tateiwa2001} or an undetectable
one ($ZrZn_{2}$)\cite{pfleiderer2001}.  It also shows
a very linear temperature dependence at low T.  This is significant in our
model because small jumps can be explained by gaplessness, which we have with the
$\beta$-fermion.  Analytically, the specific heat jump can be calculated
readily to give,
\begin{eqnarray}
\nonumber
\Delta C =\frac{\beta}{T}\frac{d\Delta^{2}}{d\beta}N(0)\bigg[
1&+&\frac{\beta}{16}JM\int\frac{d\varepsilon}{\varepsilon}\bigg[
sech^{2}\frac{\beta}{2}\left(\frac{JM}{2}+\varepsilon\right)  \\
  &-&sech^{2}\frac{\beta}{2}\left(\frac{JM}{2}-\varepsilon\right)\bigg]\bigg].
\end{eqnarray}
\noindent The first term on the right
gives the usual BCS specific heat jump, however the remaining quantity, which
is negative, gives the decrease from the BCS value. Thus,
the jump can be tuned depending on the value of the magnetization.
In Fig.\ref{spec-heat-graph} is
the numerical plot of the specific heat jump at the superconducting transition
inside the coexistent state. The specific heat is calculated with a value of $g=1.1$
which corresponds to 'strong' superconductivity, i.e., where $T_{c}{}_{\Delta}$
is large.  This can be seen in the phase diagram, Fig.\ref{phase-diagram-graph}.
Also shown for comparison are some experimental data points for $UGe_{2}$
and it is clear that the numerical calculations model well
the qualitative behavior of the specific heat. In that experiment, ref.\cite{tateiwa2001},
the specific heat is measured at 1.13 GPa where superconductivity
is maximum.  

%\section{\label{sect-discussion} Discussion}

In this paper, a mean field solution of s-wave pairing
was investigated and shown to give nontrivial results
and can be the starting point for more detailed calculations beyond mean field.
It is important to realize that the question of the nature of the
superconducting order parameter is still an open one, and we
show that a model of s-wave pairing is at least qualitatively viable.

We graciously thank J.R. Englebrecht, Y. Joglekar, N.I. Karchev, A.G. Lebed, and P.B. Littlewood
for useful discussions.
This work was done with the support of DOE Grants No. DEFG0297ER45636 and
No. 60202ER63404.

\bibliographystyle{apsrev}
\bibliography{references}
\end{document}